\documentclass[aps,twocolumn,amsmath,amssymb,superscriptaddress,pre]{revtex4}

\usepackage{graphicx}
\usepackage{dcolumn}
\usepackage{bm}
\usepackage{epsf}
\usepackage{amsmath}
\usepackage{amssymb}
\usepackage{color}

\newcommand{\bra}[1]{\langle #1|}
\newcommand{\ket}[1]{|#1\rangle}
\newcommand{\braket}[2]{\left\langle #1|#2\right\rangle}

\newcommand{\la}{\left\langle}
\newcommand{\ra}{\right\rangle}

\newcommand{\bla}{bla\\bla\\bla\\bla\\bla}

\newcommand{\PRA}{Phys. Rev. A }

\newcommand{\PRE}{Phys. Rev. E }
\newcommand{\PRX}{Phys. Rev. X }
\newcommand{\PRL}{Phys. Rev. Lett. }
\newcommand{\EPL}{EPL }

\newcommand{\NJP}{New. J. Phys. }
\newcommand{\NC}{Nat. Comm. }

\begin{document}

\title{
Performance of   shortcut-to-adiabaticity quantum engines}
\author{Obinna Abah}
\affiliation{Centre for Theoretical Atomic, Molecular and Optical Physics, Queen's University Belfast, Belfast BT7 1NN, United Kingdom}

\author{Eric Lutz}
\affiliation{Department of Physics, Friedrich-Alexander-Universit\"at Erlangen-N\"urnberg, D-91058 Erlangen, Germany}


\begin{abstract}
We consider a paradigmatic  quantum  harmonic Otto engine operating in finite time. 
We investigate its performance when shortcut-to-adiabaticity techniques are used to speed up its cycle. We   compute efficiency and power  by taking the energetic cost of the shortcut driving explicitly into account. We  analyze in detail three different shortcut methods, counterdiabatic driving, local counterdiabatic driving and inverse engineering. We demonstrate that all three lead to a simultaneous increase of efficiency and power for fast cycles, thus outperforming traditional heat engines.

\end{abstract}
\maketitle

\section{Introduction}
Heat engines have been a cornerstone of thermodynamics since the seminal work of Carnot almost 200 years ago. Carnot established that the efficiency of an engine, defined as the ratio of energy output to energy input, is maximal for quasistatic processes \cite{cal85,cen01}. Maximum efficiency is however associated with vanishing power, the rate of work production, since the quasistatic limit requires that the engine cycle is completed in an infinitely long time. For practical purposes, heat engines operate in finite time at   finite power \cite{and84,and11}.  There is generally a trade-off between  power and efficiency  in this context \cite{shi16}: increasing power leads to a decrease of efficiency, and vice versa \cite{fel00,rez06,aba12,aba16}. A current challenge is to design energy efficient thermal machines that deliver more output for the same input, without sacrificing power \cite{aps08}. 

Promising techniques to achieve this goal are collectively known as shortcuts to adiabaticity (STA). STA protocols are nonadiabatic processes  that reproduce in finite time the same final state as that of an infinitely slow adiabatic process \cite{tor13,def14}. These methods have been successfully demonstrated on a large number of experimental platforms. Examples include high-fidelity driving of a BEC \cite{bas12}, fast transport of trapped ions \cite{bow12,wal12,an16}, fast  adiabatic passage using a single spin in diamond \cite{zha13} and cold atoms \cite{du16}, as well as swift equilibration of a Brownian particle \cite{mar16}.  Different approaches to STA have been developed, such as counterdiabatic driving (CD), where a global term is added to the system Hamiltonian  to compensate for nonadiabatic transitions \cite{dem03,dem05,ber09}, local counterdiabatic driving (LCD), where the counterdiabatic term is mapped onto a local potential \cite{iba12,cam13}, and inverse engineering (IE) based on the use of dynamical invariants \cite{che10,che10a} (see Ref.~\cite{tor13} for a review).

Shortcut-to-adiabaticity methods have lately been employed to enhance the performance of classical and quantum heat engines, by reducing  irreversible losses that suppress efficiency and power \cite{den13,tu14,cam14,bea16,jar16,cho16}. However, the energetic cost of the STA driving \cite{san16,zhe16,cou16,cam16,fun16,tor17} has not been taken into account in these studies. We have recently computed efficiency and power of a quantum harmonic Otto engine, by properly including these costs, defined as the time average of the expectation value of STA term, for the case of local counterdiabatic driving (LCD) \cite{aba17}. We have found that LCD allows to simultaneously increase  efficiency and power for fast engine cycles, thus leading to  energy efficient quantum thermal machines.
   

In this paper, we extend this previous investigation to two other STA methods, namely counterdiabatic driving (CD) and inverse engineering (IE), and compare their respective  capabilities.  We specifically compute efficiency and power of  a STA quantum Otto heat engine cycle whose working medium is a time-dependent harmonic oscillator, a paradigmatic model for a quantum thermal machine \cite{rez06,aba12}.  For each STA protocol, we explicitly evaluate the cost of the STA driving  for compression and expansion phases of the engine cycle. We find that all three STA methods allow to increase, at the same time, efficiency and power  for fast cycles. We additionally show that the IE approach outperforms both CD and LCD, as it results in the largest   efficiency/power enhancement. 


\section{Quantum Otto engine}
\begin{figure}[t]
\includegraphics[width=0.95 \linewidth]{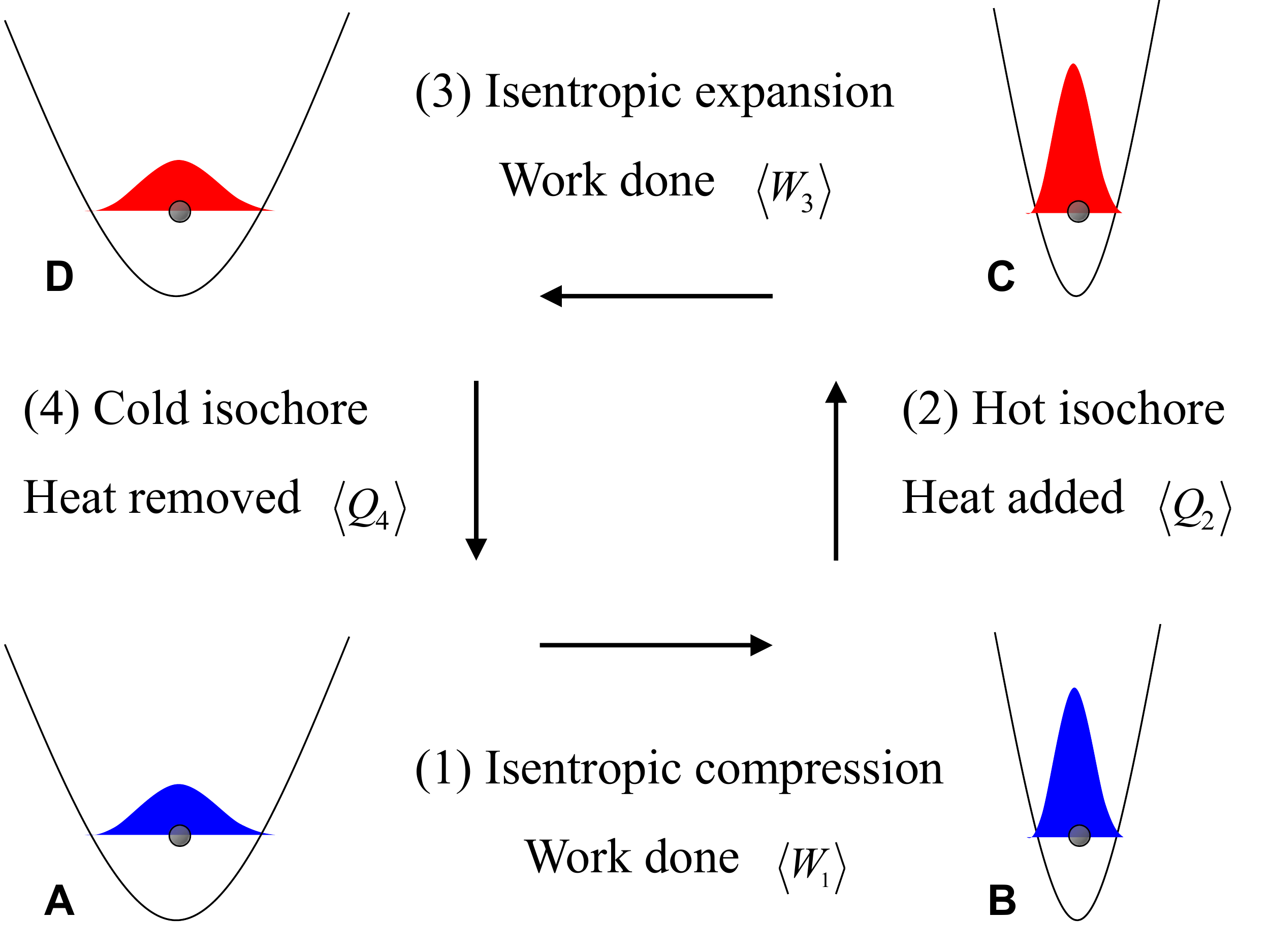}
\caption{Quantum Otto cycle of a harmonic oscillator with time-dependent  frequency. The thermodynamic cycle consists of two unitary  (compression and expansion steps 1 and 3) and two isochoric processes (heating and cooling steps 2 and 3). }
\label{fig1}
\end{figure}

We consider an Otto cycle for a time-dependent quantum harmonic oscillator.
 The corresponding Hamiltonian is of the standard  form, $H_0(t) = p^2/(2 m) + m\omega_t^2 x^2/2$, where $x$ and $p$ are the position and momentum operators of an oscillator of mass $m$. 
 As shown in Fig.~\ref{fig1}, the cycle is made of the following steps: (i) an isentropic compression branch ($AB$) where the oscillator is isolated and its  frequency $\omega_t$ is unitarily  increased  from $\omega_1$ to $\omega_2$ in a time $\tau_1$; (ii) a hot isochoric branch ($BC$) where heat is transferred from the hot bath at inverse temperature $\beta_2$ to the oscillator in a time $\tau_2$ at fixed frequency; (iii) an isentropic expansion branch ($CD$) where the frequency is modulated  to decrease  from  $\omega_2$ to $\omega_1$ in  a  time $\tau_3$;  and (iv) a cold isochoric branch ($D A$) where heat is transferred from the oscillator to the cold bath at inverse temperature $\beta_1 > \beta_2$  in a time $\tau_4$. The frequency is again kept constant. The control parameters are the time allocations on the different branches, the temperatures of the baths, and the extreme values of the modulated frequency. We will assume, as commonly done \cite{kos84,gev92,lin03,rez06,qua07,aba12,kos17}, that the thermalization times $\tau_{2,4}$ are much shorter than the compression/expansion times $\tau_{1,3}$. The total cycle time is then  $\tau_\text{cycle}= \tau_1+ \tau_3=2 \tau$ for equal step duration.

During the first and third strokes (compression and expansion), the  quantum oscillator is isolated and only work is performed by changing the frequency in time. Since the dynamic is unitary, the Schr\"odinger equation for the parametric harmonic oscillator can be solved exactly for any given frequency modulation \cite{def08,def10}. The corresponding work values are  given by \cite{aba12},
\begin{eqnarray}
\la W_1\ra &=& \frac{\hbar}{2} (\omega_2Q^\ast_1 - \omega_1) \coth\left(\frac{\beta_1\hbar\omega_1}{2}\right),\\
\la W_3\ra &=& \frac{\hbar}{2} (\omega_1 Q^\ast_3 - \omega_2) \coth\left(\frac{\beta_2\hbar\omega_2}{2}\right),
\end{eqnarray}
where we have introduced the dimensionless adiabaticity parameter $Q^*_{i}$ $(i=1,3)$ \cite{hus53}. It is defined as the ratio of the mean energy and the corresponding adiabatic mean energy and is thus equal to one for adiabatic processes \cite{def10}. Its explicit expression for any frequency modulation $\omega_t$ may be found in Refs.~\cite{def08,def10}.   On the other hand, the heat exchanged with the reservoirs during the thermalization step (the hot isochoric process) reads,
 \begin{equation}
\la Q_2\ra = \frac{\hbar \omega_2}{2} \left[\coth\left(\frac{\beta_2\hbar\omega_2}{2}\right) - Q^\ast_1 \coth\left(\frac{\beta_1 \hbar \omega_1}{2}\right) \right].
\end{equation}
For an engine,  the produced work  is negative, $\la W_1\ra +\la W_3\ra <0$, and the absorbed heat is positive, $\la Q_2\ra >0$.

The dynamics of the quantum Otto engine may be sped up with the help of STA techniques applied to the compression and expansion steps.
The STA protocols suppress the unwanted nonadiabatic transitions and thereby reduce the associated entropy production \cite{fel00,rez06,aba12,aba16}. The effective Hamiltonian of the oscillator is  then of the form,
\begin{equation}
H_\mathrm{eff}(t) = H_0(t) + H_\text{STA}^i(t),
\end{equation}
where $H_\text{STA}^i(t)$ is the STA driving Hamiltonian and $i =(1,3)$ indicates the respective compression/expansion step. The STA protocol satisfies boundary conditions  which  ensure that  initial and final expectation values  $\langle H_\text{STA}^i(0,\tau)\rangle$ vanish:
\begin{equation}
\label{5}
\begin{array}{lcr}
\omega(0) = \omega_i,& \dot{\omega}(0) = 0 ,& \ddot{\omega}(0) = 0,\\
\omega(\tau) =\omega_f,& \dot{\omega}(\tau) = 0 ,& \ddot{\omega}(\tau) = 0,
\end{array}
\end{equation}
where $\omega_{i,f}= \omega_{1,2}$ denote the respective initial  and final frequencies of  the compression/expansion steps. The  conditions \eqref{5} are, for example,  satisfied by \cite{iba12,cam13,def14},
\begin{equation}
\omega(t)= \omega_i + 10(\omega_f - \omega_i)s^3 - 15(\omega_f-\omega_i)s^4+ 6(\omega_f - \omega_i) s^5,
\end{equation}
where we have introduced  $s = t/\tau$. 

Efficiency and power are the two main quantities characterizing the performance of  a heat engine. We define the efficiency of a  STA engine as \cite{aba17},
\begin{equation}
\eta_\text{STA} = \frac{\mathrm{energy \,output}}{\mathrm{energy\, input}} = \frac{-(\la W_1\ra_\text{STA}+\la W_3\ra_\text{STA})}{\la Q_2\ra + \la H^1_\mathrm{STA}\ra_\tau + \la H^3_\mathrm{STA}\ra_\tau}, \label{6}
\end{equation} 
where $\la H_\text{STA}^i\ra_\tau = (1/\tau) \int_0^\tau dt \la H_\text{STA}^i(t)\ra$ is the time-average of the mean STA driving.
 Equation \eqref{6} takes the energetic cost of the STA driving  along the compression/expansion steps  into account. It
reduces to the adiabatic efficiency $\eta_\text{AD}$ in the absence of these two contributions.  For further reference, we additionally introduce the usual nonadiabatic efficiency of the engine, $\eta_\text{NA} = -(\la W_1\ra+\la W_3\ra)/\la Q_2\ra$,  based on the formulas (2)-(4) in the absence of any STA protocol.
 
 The power of the STA machine is on the other hand,
\begin{equation}
P_\text{STA} = -\frac{\la W_{1}\ra_\text{STA} + \la W_{3}\ra_\text{STA}}{\tau_\text{cycle}}. \label{7}
\end{equation}
Since the STA protocol ensures adiabatic work output, $\la W_{i}\ra_\text{STA} = \la W_{i}\ra_\text{AD}$ $(i=1,3)$, in a shorter cycle duration $\tau_\text{cycle}$, the superadiabatic power $P_\text{STA}$ is always greater than the nonadiabatic power $P_\text{NA}=-(\la W_{1}\ra + \la W_{3}\ra)/{\tau_\text{cycle}}$ \cite{aba17}. This ability to considerably enhance the power of a thermal machine is a key advantage of  the STA approach. In the following, we explicitly evaluate the energetic cost of the STA driving, the efficiency \eqref{6} and the power \eqref{7} for the CD, LCD and IE methods.

\section{Counterdiabatic driving (CD)} 
\begin{figure}[t]
\includegraphics[width=0.95\linewidth]{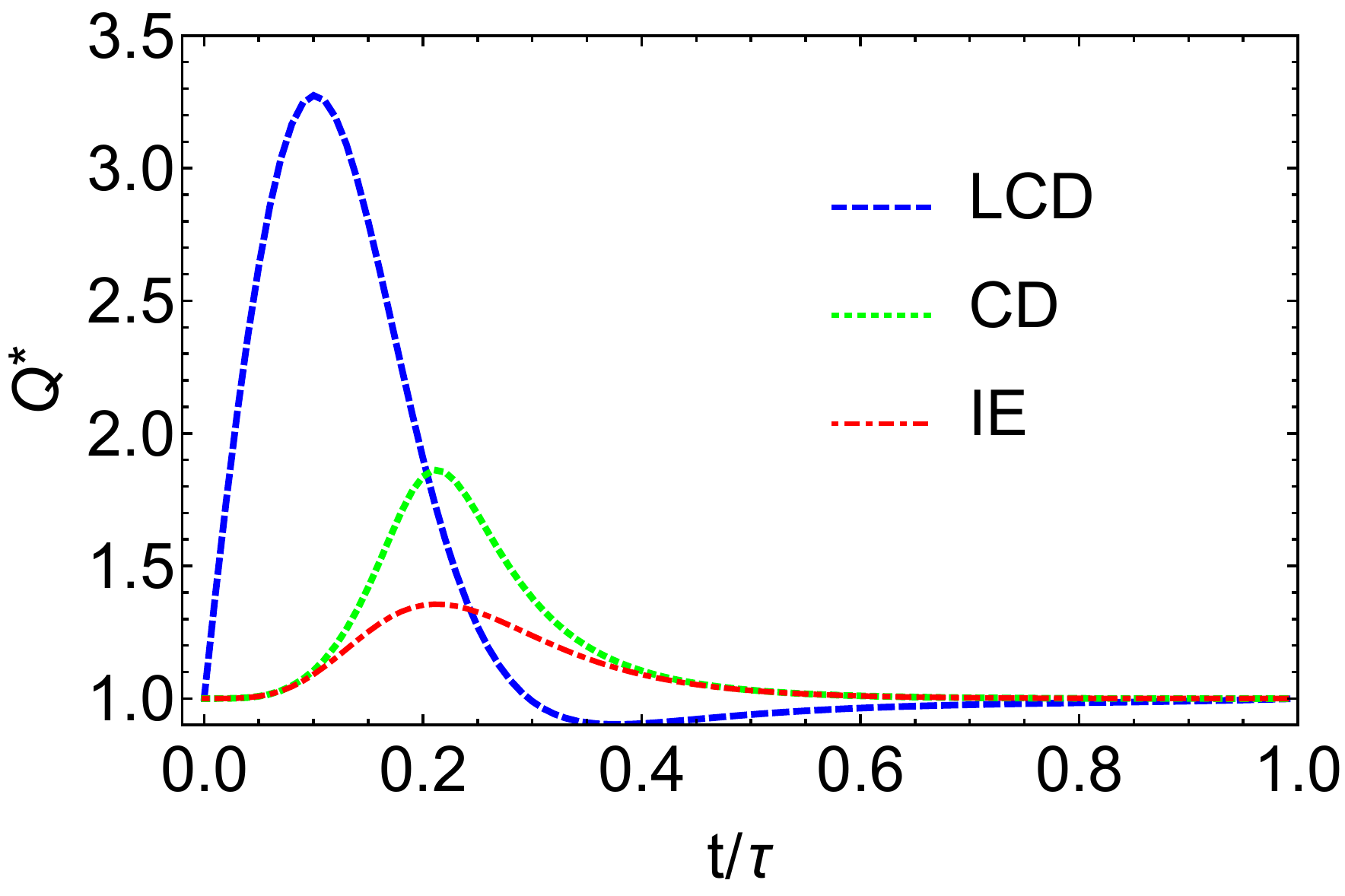}
\caption{Adiabaticity parameter $Q^*$ for the  compression  step   as a function of time  for the three shortcut methods:   counterdiabatic driving (CD) Eq.~(15) (green dotted), local counterdiabatic driving (LCD) Eq.~(21)  (blue dashed)  and   inverse engineering  (IE) Eq.~(30) (red dotted-dashed) ($\omega_1/\omega_2 = 0.15$).}
\label{fig2}
\end{figure}

We begin by analyzing the case  of counterdiabatic driving (CD),  which was first introduced by Demirplak and Rice \cite{dem03} and later independently developed by Berry \cite{ber09}. The method has recently been implemented experimentally in a trapped-ion system \cite{an16}.
The goal of counterdiabatic driving (also called transitionless quantum driving) is to find a Hamiltonian $H_\text{CD}$ for which the adiabatic approximation to the original Hamiltonian $H_0$ is the exact solution of the time-dependent Schr\"odinger equation for $H_\text{CD}$. The explicit  form of $H_\text{CD}$ is,
\begin{eqnarray}
\label{9}
H_\text{CD}(t) &=& H_0(t) + i\hbar \sum_n(\ket{\partial_t n}\bra{n} - \braket{n}{\partial_t n} \ket{n}\bra{n}) \nonumber \\
&=& H_0(t) + H_\text{STA}^\text{CD}(t),
\end{eqnarray}
where $H_\text{STA}^\text{CD}$ is the STA driving Hamiltonian. For a   time-dependent harmonic oscillator, it is given by  \cite{mug10,tor13},
\begin{equation}
H_\text{SA}^\text{CD}(t) = -\frac{\dot{\omega_t}}{4\, \omega_t}({x}{p}+ {p} {x}).
\end{equation}
The Hamiltonian \eqref{9} is quadratic in ${x}$ and ${p}$, so it may be considered describing   a generalized harmonic oscillator with a nonlocal operator \cite{ber85,mug10,che10}:
\begin{equation}
H_\text{CD}(t) = \frac{p^2}{2 m} + \left(-\frac{\dot{\omega_t}}{4 \omega_t}\right)(xp +px) + \frac{m\omega_t^2 x^2}{2}.
\label{10}
\end{equation}
Following Ref.~\cite{mis17}, we may rewrite Eq.~\eqref{10}   as, 
\begin{equation}
\label{12}
H_\text{CD} (t) = \hbar \Omega_t \left({b_t}^\dagger {b}_t + 1/2\right)
\end{equation}
with the  instantaneous ladder operators ${b}_t$,
\begin{equation}
{b}_t = \sqrt{\frac{m \Omega_t}{2 \hbar}} \left(\zeta_t {x} + \frac{i {p}}{m\Omega_t}\right),
\end{equation}
and the effective frequency,
\begin{equation}
\Omega_t = \omega_t \sqrt{1- \dot{\omega}_t^2/(4 \omega_t^4)},
\end{equation}
with $\zeta_t = 1 + \dot{\omega}_t/(2 i \omega_t \Omega_t)$.  Note that $\Omega_t^2 > 0$ to avoid trap inversion. This condition limits the rate of the frequency variation $\dot \omega_t$. Using the above equations,  the adiabaticity parameter may be simply expressed as the ratio \cite{mis17}, 
\begin{equation}
Q^\ast_\mathrm{CD} (t)= \frac{\omega_t}{\Omega_t}.
\end{equation}
The  adiabaticity parameter $Q^\ast_\mathrm{CD}$ is plotted as a function of the  time $t/\tau$  for the compression step  in Fig.~\ref{fig2} (the corresponding result for the expansion  is simply the mirror image). We observe that $Q^\ast_\mathrm{CD}$ approaches the adiabatic value one at the end of the driving, as it should, and it is much smaller than the nonadiabatic  $Q^\ast_\mathrm{NA}$, as expected.

\begin{figure}[t]
\includegraphics[width=0.95 \linewidth]{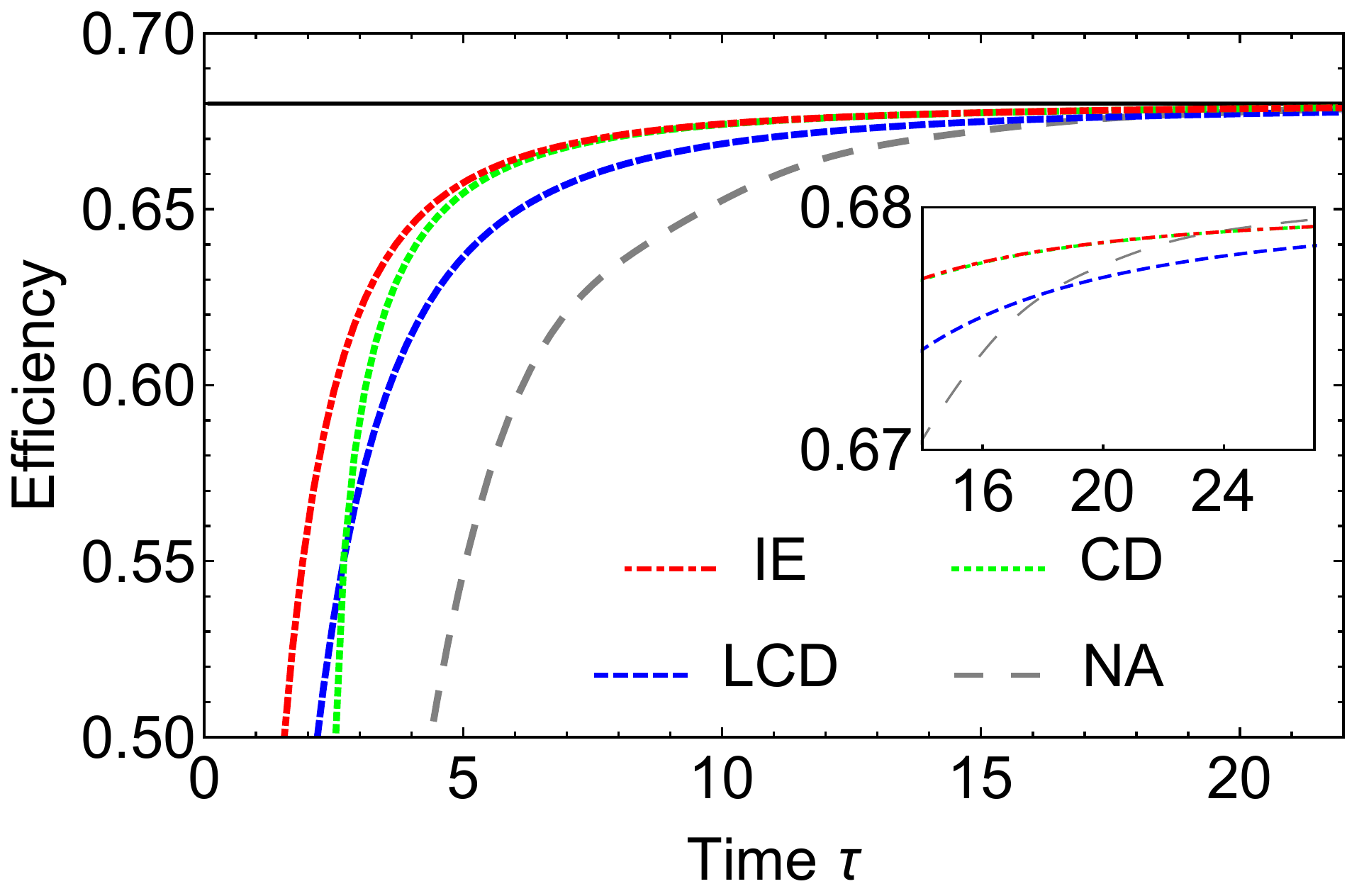}
\caption{Efficiency  as a function of the driving time $\tau$ for the three shortcut methods: counterdiabatic driving (CD)  (green dotted), local counterdiabatic driving (LCD)   (blue dashed)  and   inverse engineering  (IE)  (red dotted-dashed). The gray (large dashing) line shows the nonadiabatic efficiency (NA) without shortcut, while the black (solid) horizontal line is the adiabatic efficiency. Parameters are $\omega_1 = 0.32$, $\omega_2 = 1$, $\beta_1 =  0.5$ and $\beta_2 = 0.05$. }
\label{fig3}
\end{figure}

We proceed by evaluating the mean energy of the effective harmonic oscillator \eqref{12} at any time $t$, assuming that it is initially in a  thermal state $P_n^i = \exp (-\beta_i E_n^i)/Z_i$ at inverse temperature $\beta_i$.  We obtain,
\begin{eqnarray}
\langle H_\mathrm{CD}(t) \rangle &=& \sum_{n} \hbar \omega_t (\langle m \rangle_{n,t} + 1/2) P_n^i \nonumber\\
&=&  \frac{\omega_t}{\omega_i}  Q^\ast_\mathrm{CD}\langle H(0)\rangle,
\end{eqnarray}
where we have used the following expression for the mean quantum number,
$\langle m \rangle_{n,t} + 1/2 = (n +1/2) Q^\ast_\mathrm{CD}$ \cite{def08,def10}, and $\langle H(0)\rangle = \hbar \omega_i\coth(\beta \hbar \omega_i/2)/2$. 
The expectation value of the CD driving finally follows as,
\begin{equation}
\langle H_\mathrm{STA}^\mathrm{CD}\rangle = \langle H_\mathrm{CD}(t) \rangle - \langle H_0(t) \rangle = \frac{\omega_t}{\omega_i} \langle H(0) \rangle \left(\frac{\omega_t}{\Omega_t} - 1\right),
\label{17}
\end{equation}
where we used $\langle H_0(t)\rangle = \langle H(0)\rangle \omega_t /\omega_i$ \cite{def10}. We numerically compute the energetic cost of the STA driving as the time average of Eq.~\eqref{17}. This time average is different from zero, although $\langle H_\mathrm{STA} (t=(0, t_f))\rangle  = 0$ in view of the boundary conditions  \eqref{5}. The corresponding efficiency \eqref{6} and power \eqref{7} are shown as a function of the driving time $\tau$ in Figs.~\ref{fig3} and \ref{fig4} (green dotted). 

\section{Local counterdiabatic driving (LCD)}

\begin{figure}[t]
\includegraphics[width=0.95 \linewidth]{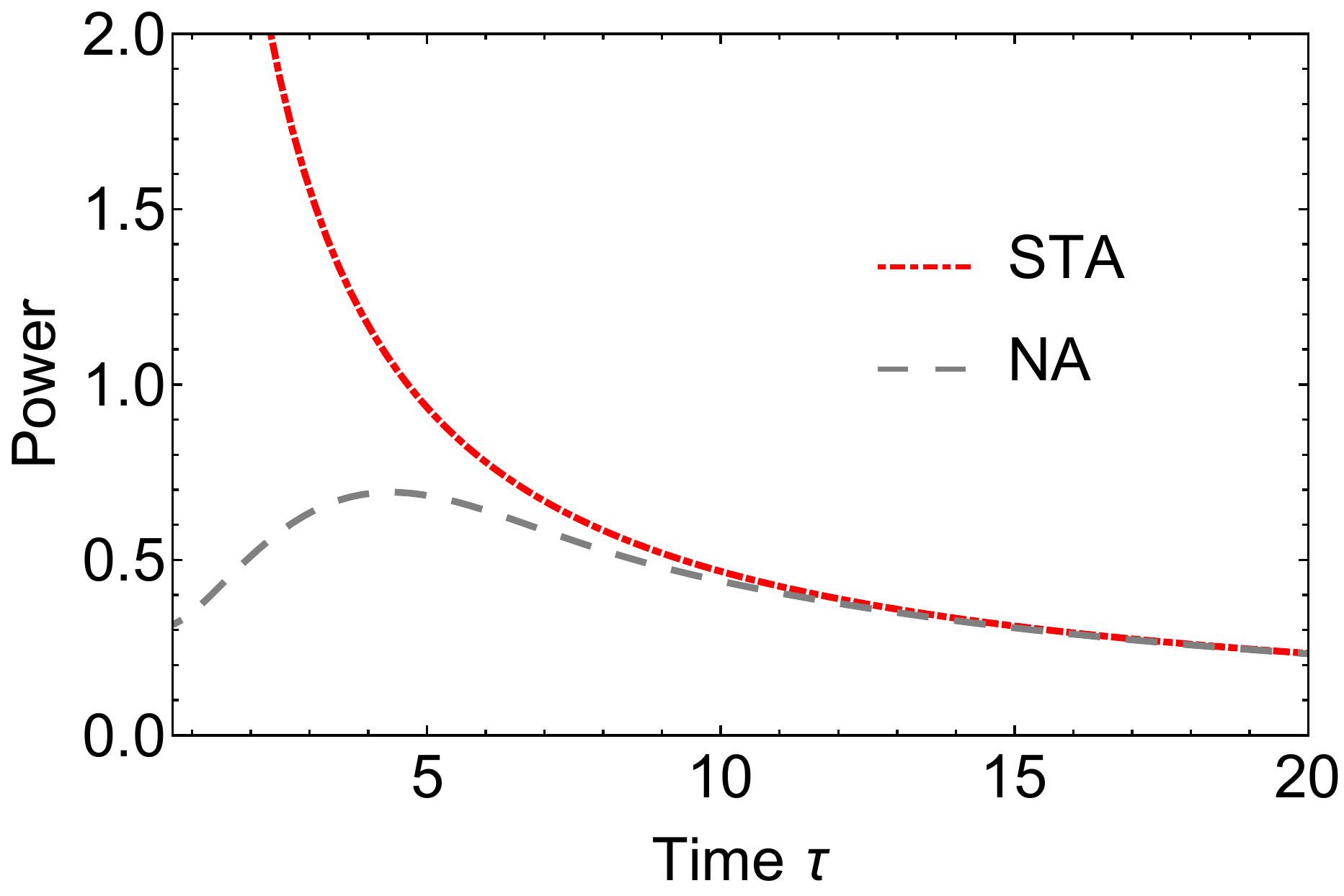}
\caption{Power  as a function of the driving time $\tau$ for the three shortcut methods: CD, LCD and IE lead to the same power  (red dotted-dashed). The gray (large dashing) line shows the nonadiabatic (NA) efficiency.  Same parameters as in  Fig.~3}
\label{fig4}
\end{figure}

 A limitation of the CD method is that it  requires the knowledge of the spectral properties of the original Hamiltonian $H_0(t)$ at all times to construct the auxiliary term $H_\text{STA}^\text{CD}(t)$ in Eq.~\eqref{9}. A possibility to circumvent this problem is offered by the local counterdiabatic (LCD) approach \cite{iba12,cam13}, which has been experimentally demonstrated in Refs.~\cite{sch10,sch11,bas12}.  Here the nonlocal operator (\ref{10}) is mapped onto a unitarily equivalent Hamiltonian with a local   potential by applying the canonical transformation, $U_x = \exp\left({i m \dot{\omega}_t x^2}/{4 \hbar \omega}\right)$, which cancels the cross terms $xp$ and $px$. This leads to a new local counterdiabatic (LCD) Hamiltonian of the form \cite{iba12,cam13},
\begin{eqnarray}
H_\mathrm{LCD}(t) &=& U_x^\dagger (H_\mathrm{CD}(t) - i\hbar \dot{U}_x U_x^\dagger) U_x \nonumber \\ &=& \frac{p^2}{2 m} + \frac{m\tilde\Omega_t^2 x^2}{2},
\label{19}
\end{eqnarray}
with the modified time-dependent squared frequency,
\begin{equation}
{\tilde \Omega}^2(t) = \omega_t^2 -\frac{3\dot{\omega}_t^2}{4\omega_t^2}+\frac{\ddot{\omega}_t}{2\omega_t}. 
\label{20}
\end{equation}
The Hamiltonian \eqref{19} still drives the evolution along the adiabatic trajectory of the system of interest. By demanding that $H_\mathrm{LCD} (0,\tau)= H_\mathrm{0}(0,\tau)$, and imposing $\dot{\omega} (\tau) = \ddot{\omega} (\tau) = 0$, the final state is  equal for both dynamics, even in phase, and the final vibrational state populations coincide with those of a slow adiabatic process \cite{iba12}. The frequency $\tilde\Omega^2(t)$ approaches $\omega^2(t)$ for very slow expansion/compression  \cite{cam13}. The LCD technique may be applied as long as $\tilde \Omega_t^2 > 0$.
The expectation value of  the local counterdiabatic Hamiltonian may be  computed in analogy to the conterdiabatic driving and reads \cite{aba17}, 
\begin{eqnarray}
\langle H_\mathrm{LCD}(t) \rangle &=& \frac{\omega_t}{\omega_0} \left(1 - \frac{\dot{\omega}_t^2}{4\omega_t^4} + \frac{\ddot{\omega}_t}{4 \omega_t^3} \right) \langle H(0) \rangle \nonumber\\
 &=& \frac{\omega_t}{\omega_0} Q_\mathrm{LCD}^\ast \langle H(0)\rangle,
\end{eqnarray}
with the adiabaticity parameter,
\begin{equation}
Q^\ast_\text{LCD} (t) = 1 - \frac{\dot{\omega}_t^2}{4 \omega_t^4} + \frac{\ddot{\omega}_t}{4\omega_t^3}.
\end{equation}
The variation of the adiabaticity parameter as a function of $t/\tau$ is shown in Fig.~\ref{fig2}. The expectation value of the LCD driving  is moreover evaluated as before \cite{aba17},
\begin{eqnarray}
\langle H_\mathrm{STA}^\mathrm{LCD} (t)\rangle &=& \langle H_\mathrm{LCD} (t)\rangle - \langle H_0(t)\rangle \nonumber \\
&=& \frac{\omega_t}{\omega_0}\langle H(0)\rangle \left[- \frac{\dot{\omega}_t^2}{4 \omega_t^4} + \frac{\ddot{\omega}_t}{4 \omega_t^3}\right].
\end{eqnarray}
The corresponding numerically computed efficiency \eqref{6} and power \eqref{7} are shown as a function of the driving time $\tau$ in Figs.~\ref{fig3} and \ref{fig4} (blue dashed).

\section{Inverse engineering (IE)}
An additional STA method is based on the design of  appropriate parameter trajectories of the frequency by employing the  Lewis-Riesenfeld invariants of motion \cite{lew69} supplemented by simple inverse-problem techniques \cite{pal98}.
For the time-dependent harmonic oscillator described by  $H_0(t)$, the state dynamics will be the solution of the corresponding Schr\"odinger equation based on the invariants of motion of the following form \cite{che10,che10a},
\begin{equation}
I(t) = \frac{1}{2}\left(\frac{{x}^2}{\bar b^2}m \omega_0^2 + \frac{1}{m}{\pi}^2\right),
\end{equation}
where ${\pi} = \bar b{p} - m\dot{\bar b} {x}$ plays the role of a momentum conjugate to ${x}/\bar b$  and $\omega_0$ is in principle an arbitrary constant. The 
dimensionless scaling function $\bar b_t = \bar b(t)$ satisfies the Ermakov equation, 
\begin{equation}
\ddot {\bar b}_t + \omega^2_t \bar b_t = \omega_0^2/\bar b_t^3.
\label{23}
\end{equation}
Its solutions should be chosen real 
to make ${I}$ Hermitian. Whereas $\omega_0$ is often rescaled to unity
by a scale transformation of $\bar b_t$, another convenient choice is $\omega_0 = \omega(0)$.
To achieve STA processes,  $\omega(t)$ is first left undetermined and $\bar b_t$ is set to fulfil the equations ${I}(0) = H_0(0)$ and $[{I}(t_f),{H}_0(t_f)]=0$. This guarantees that the eigenstates of ${I}$ and ${H}_0$ are the same at the initial and final times and can be done by satisfying the boundary conditions,
\begin{equation}
\begin{array}{lcr}
\bar b(0) = 1,& \dot{\bar b}(0) = 0 ,& \ddot{\bar b}(0) = 0,\\
\bar b(\tau) =\sqrt{\omega_0/\omega_f}= \gamma,& \dot{\bar b}(\tau) = 0 ,& \ddot{\bar b}(\tau) = 0,
\end{array}
\end{equation}
with $\omega_0 = \omega(0)$ and $\omega_f = \omega(\tau)$. For an individual eigenstate $n$ of the oscillator Hamiltonian, the corresponding time-dependent instantaneous energy is, 
\begin{equation}
\la H_\mathrm{IE}(t)\ra_n = \frac{\hbar (n +1/2)}{2\omega_0}\left(\dot{\bar b}_t^2 + \omega^2_t \bar b_t^2+\frac{\omega_0^2}{\bar b^2_t}\right).
\end{equation}
The parameter $\omega_t$ is here deduced from the Emarkov equation \eqref{23}. To  ensure the non-inversion of the trap, the condition $t_f > 1/(2\omega_f)$ should be  satisfied. The expectation value of the STA  at any given time follows as \cite{che10,che10a},  
\begin{equation}
\langle H_\mathrm{IE} (t)\rangle = \frac{\hbar}{2} \left[\frac{\dot{\bar b}_t^2}{2\omega_0} + \frac{\omega_t^2 \bar b_t^2}{2 \omega_0} + \frac{\omega_0}{2 \bar b_t}\right] \coth\left(\frac{\beta \hbar \omega_0}{2}\right).
\label{28}
\end{equation}
Using the relation $\bar b_t = (\omega_0/\omega_t)^{1/2} $, we further have, 
\begin{equation}
\dot{\bar b}_t = - \frac{1}{2}\left(\frac{\omega_t}{\omega_0}\right)^{1/2} \frac{\omega_0 \dot{\omega}_t}{\omega_t^2}  \hspace{0.5cm} \text{and}  \hspace{0.5cm}
\dot{\bar b}_t^2 = \frac{1}{4}\frac{\omega_0 \dot{\omega}_t^2}{\omega_t^3}.
\label{29}
\end{equation}
Combining Eqs.~\eqref{28} and \eqref{29}, the time-dependent expectation value \eqref{28} can finally be written as, 
\begin{eqnarray}
\langle H_\mathrm{IE} (t)\rangle &=& \frac{\hbar}{2} \left[\frac{\dot{\omega}_t^2}{8 \omega^3} + \frac{\omega_t}{2} + \frac{\omega_t}{2}\right] \coth\left(\frac{\beta \hbar \omega_0}{2}\right)\nonumber \\
& =& \frac{\omega_t}{\omega_0} \langle H(0)\rangle \left[1 + \frac{\dot{\omega}_t^2}{8 \omega_t^4}\right].
\end{eqnarray}
The associated adiabaticity parameter hence reads,
\begin{equation}
Q_\mathrm{IE}^{\ast}(t) = 1 + \frac{\dot{\omega}_t^2}{8 \omega_t^4},
\end{equation}
as shown in Fig.~\ref{fig2} as a function of $t/\tau$.
We may again deduce the expectation value of the IE driving  as,  
\begin{equation}
\langle H_\mathrm{STA}^\mathrm{IE} (t)\rangle = \langle H_\mathrm{IE} (t)\rangle - \langle H_0(t)\rangle = \frac{\omega_t}{\omega_0}\langle H(0)\rangle \frac{\dot{\omega}_t^2}{8 \omega_t^4}.
\end{equation}
The corresponding numerically computed efficiency \eqref{6} and power \eqref{7} are shown as a function of the driving time $\tau$ in Figs.~\ref{fig3} and \ref{fig4} (red dotted-dashed).

\begin{figure}[t]
\includegraphics[width=0.95 \linewidth]{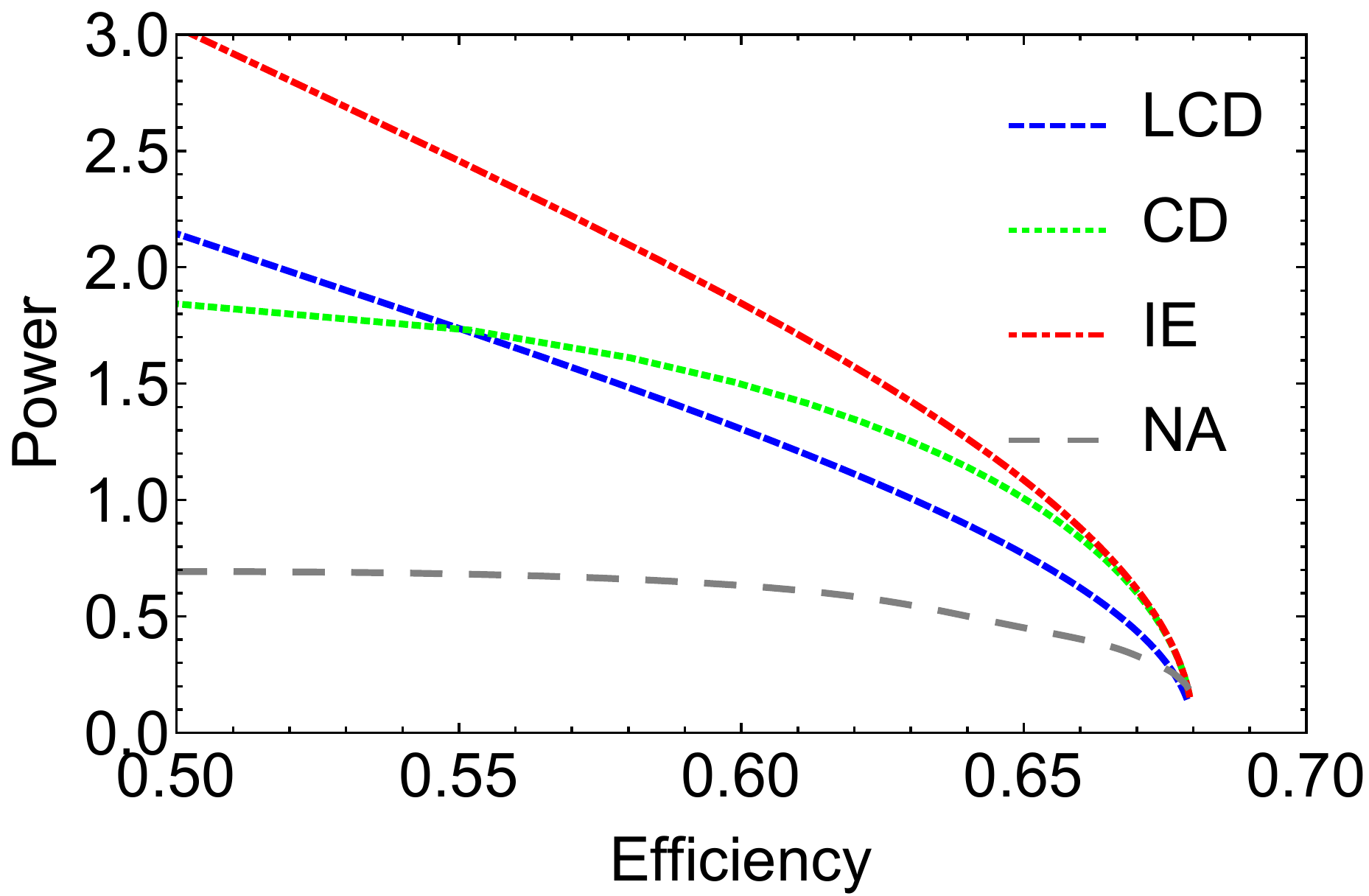}
\caption{Power-efficiency diagram for the three shortcut methods: counterdiabatic driving (CD)  (green dotted), local counterdiabatic driving (LCD)   (blue dashed)  and   inverse engineering  (IE)  (red dotted-dashed). The gray (large dashing) line shows the nonadiabatic efficiency (NA) without shortcut. Same parameters as in Fig.~3.   }
\label{fig5}
\end{figure}

\section{Discussion and conclusions}
We have performed a detailed analysis of the performance of a STA quantum harmonic heat engine, using three commonly employed techniques: CD, LCD and IE. These three methods emulate adiabatic processes in finite time. We have first compared the time-dependent adiabaticity parameter $Q^*(t)$, Eqs.~(15), (21) and (30), for all three STA approaches (shown in Fig.~\ref{fig2}). We observe that while all three methods lead to $Q^*(\tau)=1$, per construction, the time dependence of $Q^*(t)$ may widely differ. The overall lowest value is achieved by inverse engineering (IE), which therefore appears to be the most effective technique to reduce unwanted nonadiabatic transitions.

We have further numerically calculated the energetic cost of the STA driving as the time average of the expectation value of the respective STA Hamiltonians, given in Eqs.~(17), (22) and (31). The corresponding efficiencies and powers, that take into account this energetic cost, are presented in Figs.~\ref{fig3} and \ref{fig4}. We first note that all three methods lead to a significant increase of the efficiency at short cycle times, compared to the standard NA engine without shortcut. We observe furthermore that they all simultaneously yield a large enhancement of the power in the same regime. STA engines thus always outperform their traditional counterparts for short cycle durations. This is a remarkable feature of STA boosted quantum heat engines. They hence appear as energy efficient thermal machines that are able to produce more output from the same input at higher power. This  property follows from the fact that STA protocols, on the one hand,  speed up the dynamics (therefore increasing power), and on the other hand, also ensure that the final  state is an adiabatic state instead of a highly excited state (thus reducing entropy production and consequently increasing efficiency).

Our study finally establishes that among the three considered STA methods, inverse engineering (IE) offers the largest increase of efficiency. This  result  confirms and directly follows from our previous observation   that IE is the most effective method to suppress nonadiabatic transitions. At the same time, all three approaches yield  the same enhanced power, since they produce the adiabatic work output in much less time. These findings are illustrated in the power-efficiency diagram shown in Fig.~5. The latter clearly demonstrates the advantage of STA heat engines operating in finite-time.

\acknowledgments
{This work was partially supported by the EU Collaborative Project TherMiQ (Grant Agreement 618074).  OA was supported by the Royal Society Newton International Fellowship (grant number NF160966) and the Royal Commission for the 1851 Exhibition.}

\end{document}